\documentclass[conference]{IEEEtran}
\IEEEoverridecommandlockouts
\usepackage{cite}
\usepackage{amsmath,amssymb,amsfonts}
\usepackage{algorithmic}
\usepackage{algorithm}
\usepackage{graphicx}
\usepackage{array}
\usepackage{textcomp}
\usepackage{xcolor}
\usepackage{siunitx} 
\usepackage{comment}
\usepackage{threeparttable}
\def\BibTeX{{\rm B\kern-.05em{\sc i\kern-.025em b}\kern-.08em
    T\kern-.1667em\lower.7ex\hbox{E}\kern-.125emX}}

\usepackage{cleveref}    

\begin{document}

\title{NeuDW-CIM: a 65-nm 0.8-pJ/Sop Reconfigurable Neuromorphic Compute-in-Memory Macro with Nonlinear Dendrites and K-Winners\\


\author{
    \IEEEauthorblockN{
        Junyi Yang$^{1}$, 
        Yahan Yang$^{1}$,
        Shuai Dong$^{1}$,
        Biyan Zhou$^{1}$,
        Ye Ke$^{1}$,
        Zhengnan Fu$^{1}$,\\
        Xin Si$^{2}$,
        An Guo$^{1}$,
        Peng Zhou$^{1}$,
        Arindam Basu$^{1}$$^{\ast}$
    }
    \IEEEauthorblockA{
        $^{1}$Department of Electrical Engineering, City University of Hong Kong, Hong Kong, China\\
        $^{2}$Southeast University, Nanjing, China. (*Corresponding email: arinbasu@cityu.edu.hk) 
    }
}


}

\maketitle


\begin{abstract}
This work presents NeuDW-CIM, a highly efficient neuromorphic Compute-in-Memory (CIM) macro for Spiking Neural Networks (SNNs) implemented in 65 nm CMOS. The design introduces a custom twin 9T bit-cell for ternary inputs/weights and a reconfigurable non-linear In-Memory ADC (IMA). The macro supports two specialized modes: 1) Nonlinear Dendrite (NLD) mode, which utilizes reconfigurable IMA to emulate biological dendritic functions, achieving measured accuracies of 97.2\% on N-MNIST and 95.5\% on DVS Gesture; and 2) Top-K Winner (KWN) mode, featuring an early-stopping mechanism that reduces IMA conversion latency by 30\% and digital LIF latency by 10$\times$. Benefiting from the sparse update in KWN mode, NeuDW-CIM achieves a measured energy efficiency (EE) of 0.8 pJ/SOP (1.6× improvement). 
\end{abstract}



\section{Introduction}
 Neuromorphic SNNs have proven to be very energy-efficient due to their sparse, binary activations \cite{1zhang202322,ESSRCICsharma202565nm} and can provide high-speed, low-latency operation when coupled with dynamic vision sensors (DVS) \cite{3niwa20232}. Recent work \cite{4liu202430} has combined the efficiency of analog CIM for synaptic operations and event-driven clock gating for saving dynamic energy. However, using analog circuits with capacitor based leaky integrate and fire (LIF) neurons reduces reconfigurability. Digital LIF following analog CIM provides flexibility, but faces several other architectural challenges (Fig. 1 (a)). First, it requires an analog-to-digital Converter (ADC) after the analog Multiply-Accumulate (MAC), which increases power and area overheads \cite{ESSRICchoi2024ar,6kim2023neuro}. Moreover, it suffers from poor throughput due to serial membrane potential ($V_{mem}$) update of the digital LIF. Third, the output of event sensors [3] is normally ternary with ON/OFF events, which require more channels when mapped to conventional binary input CIM. Lastly, the accuracy of SNN on a limited-precision analog CIM has been limited.  
 
 To overcome these issues, a \underline{Neu}romorphic CIM with \underline{D}endrites and \underline{W}inners (NeuDW-CIM) (Fig. 1 (b)) supporting top-K winner (KWN) and Non-linear (NL) dendrites (NLD) modes is proposed with the following features: \textbf{1)  NL reconfigurable in-memory ADC (IMA) with small area overhead (12\%):} implementing different NL activations with measured average error $<$1 LSB, achieving high accuracy in NLD mode. \textbf{2) KWN with early stop:} A KWN selection with an early stop mechanism reducing IMA latency by 30\% and digital LIF latency by 10$\times$, boosting energy efficiency by 1.6$\times$. \textbf{3) Custom twin 9T bit-cell with multi-VDD:} efficiently supporting ternary inputs/weights with enhanced weight precision and minimal overhead.  \textbf{4) Accuracy optimization:} integrating NL quantization (NLQ) and a sensitive neuron list (SNL) to enhance inference accuracy in KWN mode, with silicon results showing a measured NLQ mean error of 0.41 LSB.  

\begin{figure}[t]
  \centering
  \includegraphics[width=0.95\linewidth]{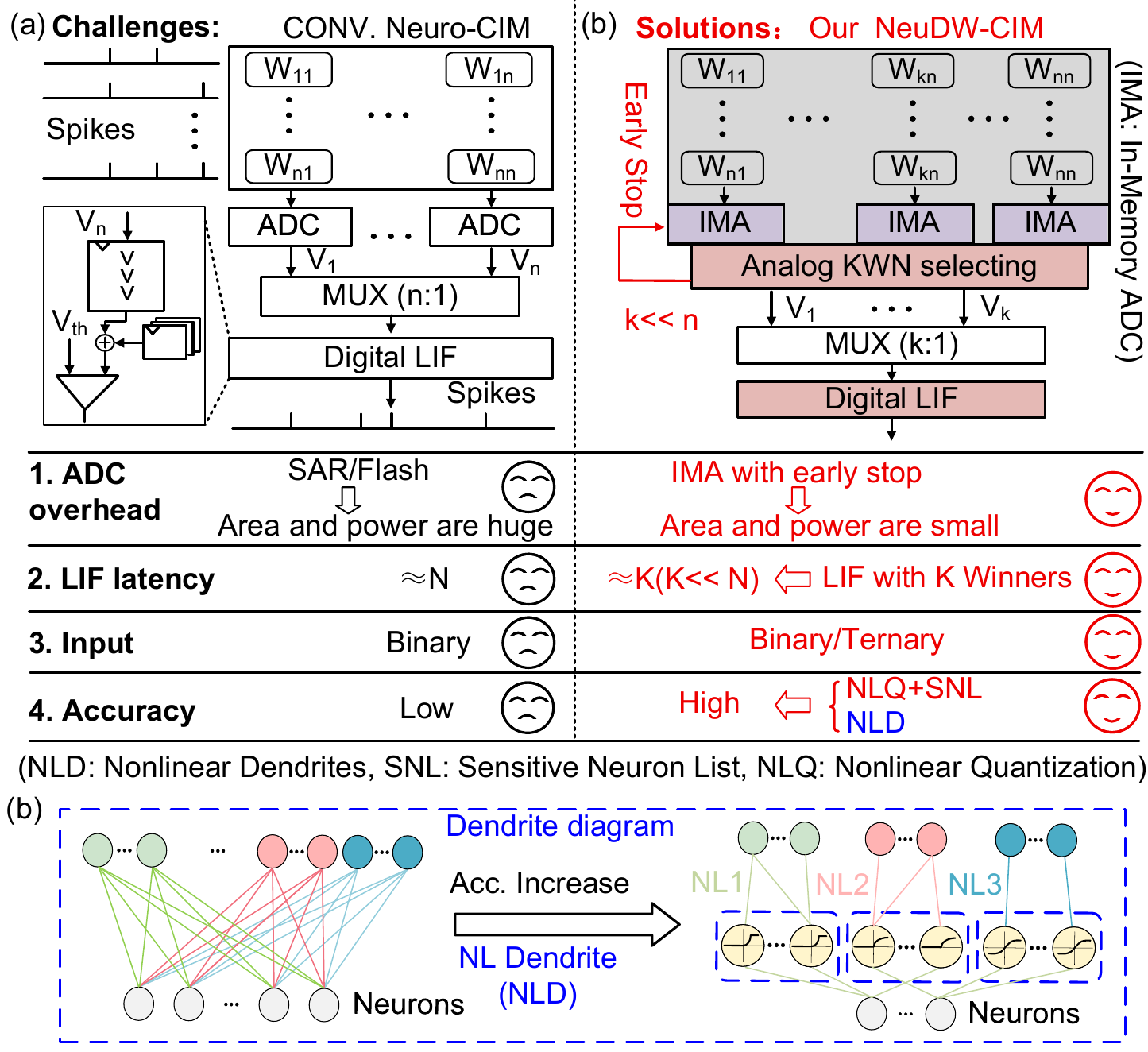}
  \caption{(a) Challenges and (b) solutions of neuromorphic CIM. (c) Dendrite for increasing accuracy.}
  \label{Challenges and solutions}
\end{figure}
\section{Proposed NeuDW-CIM Architecture}
Fig. 2 depicts the architecture of the NeuDW-CIM with two modes. It consists of 256×128 and 46×128 static random-access memory (SRAM) arrays for MAC and NL-IMA, respectively. The digital controller configures the ramp ADC for dual task of NL quantization and selection of top-K largest MAC results (K$\ll$128) in KWN mode. Only these indices are used to update $V_{mem}$ for the digital LIF, and the ramp is stopped early after the first K conversions resulting in saving ADC/LIF latency and energy.  
Equation (1) describes the $V_{mem}$ update logic in KWN mode featuring a top-K selection mechanism, where only the K winner neurons undergo membrane potential updates to effectively reduce LIF conversion latency. 
\begin{equation}
V^p_{mem}(\mathrm{t}+1)=\left\{\begin{array}{l}
\left(\sum_i W_{i, p} S_i\right)+\beta V^p_{mem}(\mathrm{t})+\mathrm{n}(\mathrm{t}) \\ \text {(p in top-K)} \\
V^p_{mem}(\mathrm{t}) \quad \text {(p not in top-K)}
\end{array}\right.
\label{eq:KWN}
\end{equation}
 In NLD mode, various biological dendritic functions with NL activations are integrated into the network to enhance accuracy as illustrated in Fig. 1(c). These NL activations are implemented by the NL-IMA, where outputs from the ramp-IMA are digitized via
a ripple counter. Owing to the inherent sparsity of the connections, this enhancement is achieved without increasing the total parameter overhead. Equation (2) defines the update logic of neurons emulating biological NL dendritic characteristics using $f$(), which enhances the learning capability of SNN neurons and improves overall system accuracy.
\begin{equation}
V^p_{mem}(\mathrm{t}+1)=\sum_j W_{j, p}^d f\left(\sum_i W_{i, j, p}^s S_i\right)+\beta V^p_{mem}(\mathrm{t})
\label{eq:NLD}
\end{equation}  
\begin{figure}[t]
  \centering
  \includegraphics[width=\linewidth]{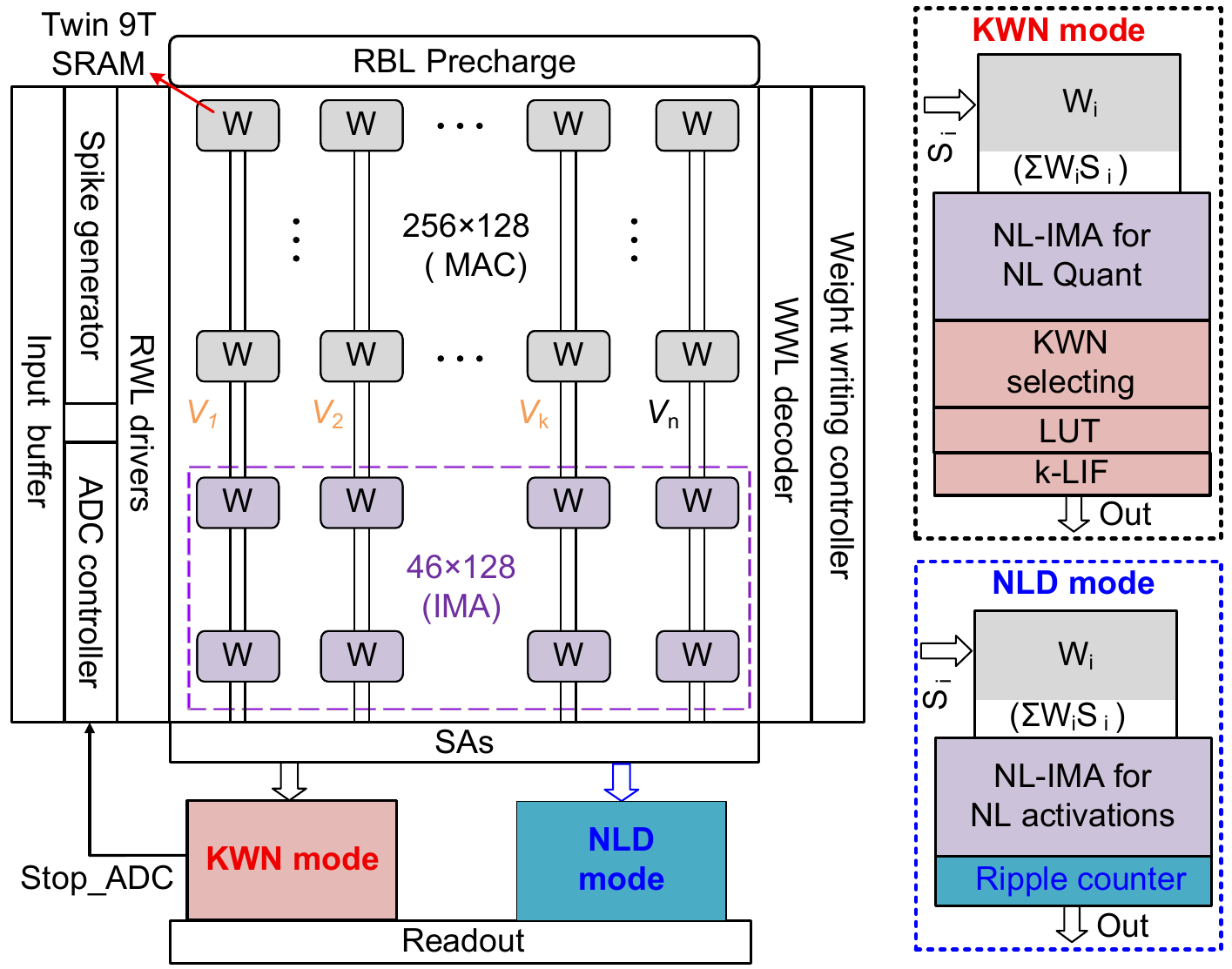}
  \caption{Overall architecture of the proposed macro and two modes.}
  \label{Overall architecture}
\end{figure}   
\subsection{Twin 9T bit-cell with Multi-VDD Array}\label{Twin 9T bitcell}
Fig. 3(a) shows the schematic of a custom twin 9T SRAM bit-cell that enables multiplication between a ternary input (represented by a pair of +/- RWL signals) and a ternary weight (represented by two 6T-SRAM bit-cells). The SRAM array is partitioned into two banks, each powered by a dedicated supply voltage: $VDD_{MSB}$ and $VDD_{LSB}$ as shown in Fig. 3(b), and the two VDDs are implemented using external power inputs, incurring a minimal power overhead of only 3.5 $\mu$W. This Multi-VDD scheme maps the stored weight’s MSB and LSB to two distinct discharge currents, $I_{MSB}$ and $I_{LSB}$, with a fixed ratio such as $I_{MSB}$ = 2$I_{LSB}$ for 3-bit weights. Monte Carlo (MC) simulations show minimal fluctuation in the current ratio ($I_{MSB}/I_{LSB}$) in Fig. 3(c). We also compare the latency and bit-cell count advantage of our method with conventional pulse-width modulation (PWM) \cite{7dong2025topkima}, and multi-cell (MCL) \cite{8yu202265} approaches for implementing multi-bit weight in Fig. 3(d), showing 4× and 7.8× advantages for the 5-bit weight case. 
\begin{figure}[t]
  \centering
  \includegraphics[width=\linewidth]{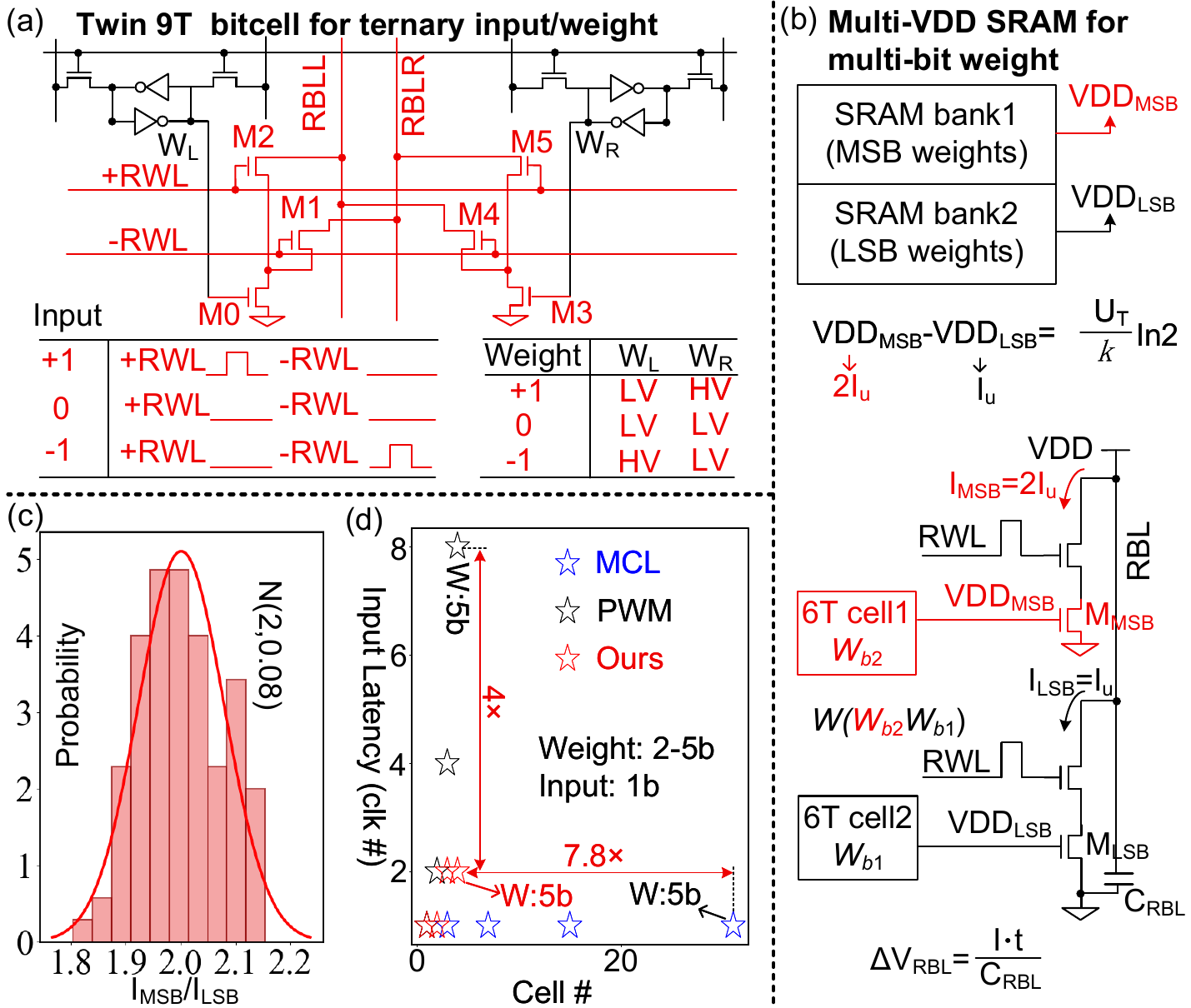}
  \caption{(a) Custom twin 9T SRAM bit-cell. (b) Multi-VDD SRAM for multi-bit weight. (c) MC simulation of the current ratio. (d) Overhead comparison for implementing multi-bit weight.}
  \label{Twin 9T bitcell and multi-Vdd SRAM}
\end{figure}
\subsection{KWN/NLD modes}
\label{modes}
\begin{figure}[t]
  \centering
  \includegraphics[width=0.95\linewidth]{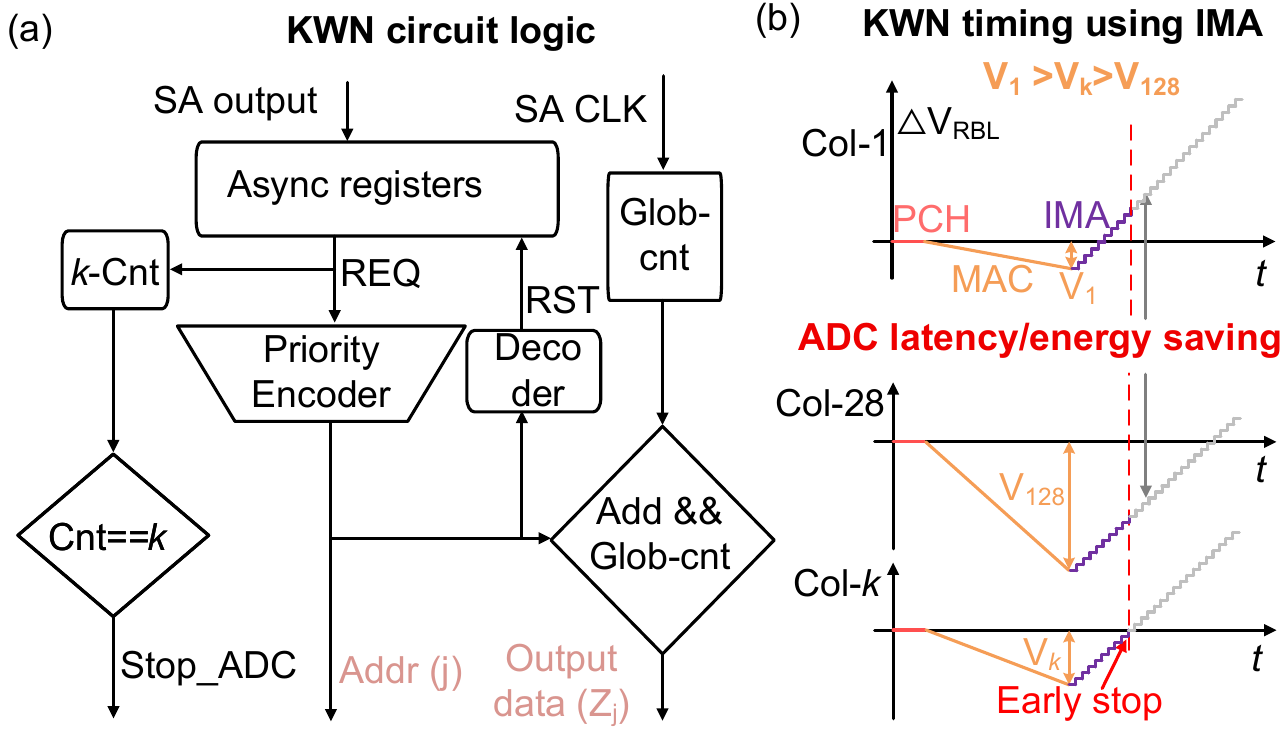}
  \caption{(a) Circuit controller and (b) timing of the KWN selecting module. }
  \label{Schematic and timing of the KWN selecting module}
\end{figure}
 Fig. 4 depicts the KWN circuit controller and IMA timing, where the IMA is used to create a differential ramp voltage on the read bitlines (RBLs) by sequentially turning on rows after an initial result of MAC is stored on the RBL. Once zero-crossings are detected on the first K RBLs, the controller stops the ADC array according to the early stopping mechanism ($Stop\_ADC$ in Fig. 4(a)). The early stopping mechanism achieves a 30\% reduction in ADC conversion latency for the DVS Gesture dataset. For each crossing, the column index, j, obtained from a priority encoder (PENC) and counter value, $Z_j$ (corresponding to quantized MAC), are stored in registers as outputs. 
\begin{figure}[t]
  \centering
  \includegraphics[width=\linewidth]{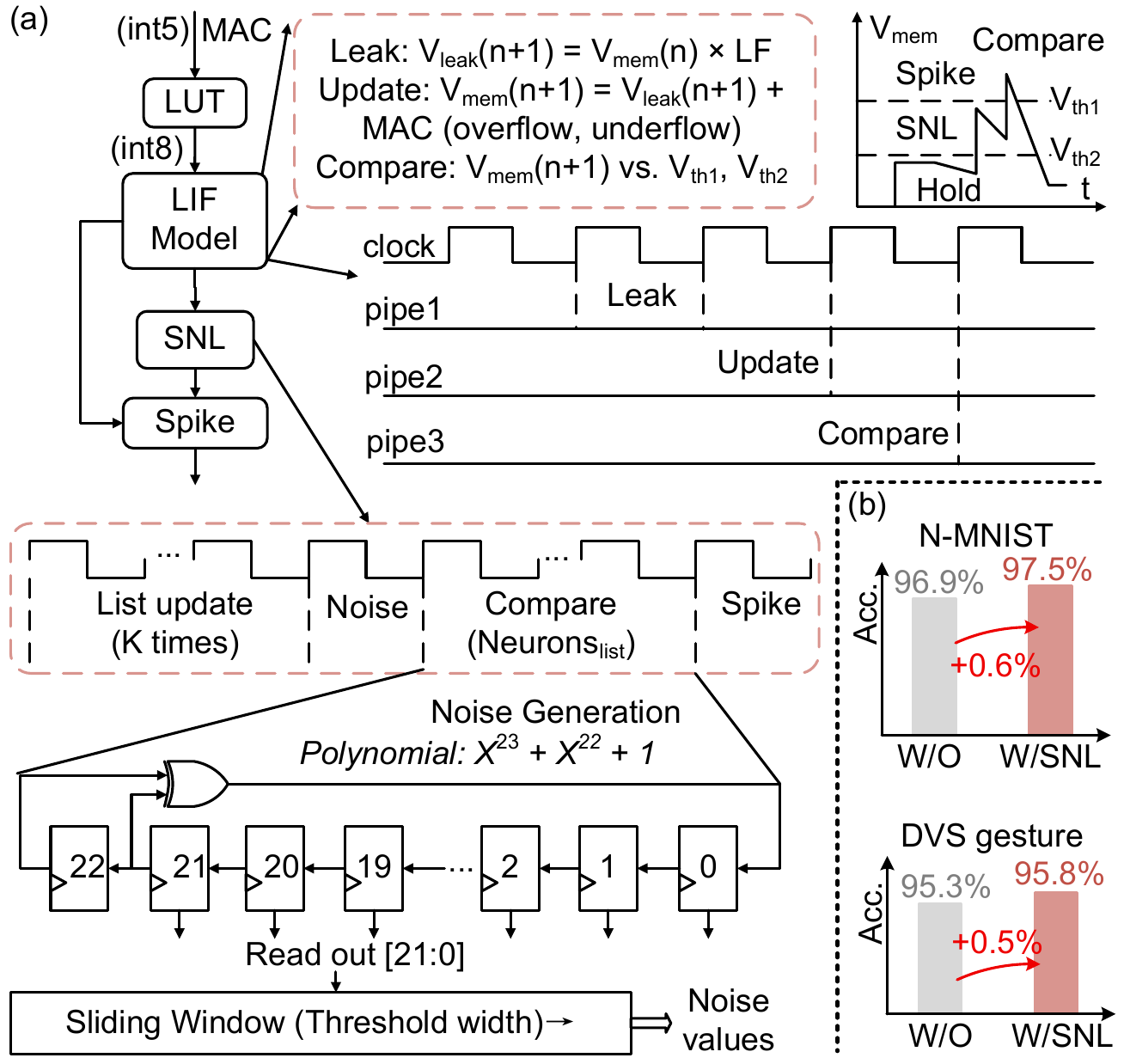}
  \caption{(a) Flow chart and timing of the LIF module with noise. (b) Accuracy improvement with SNL.}
  \label{Flow chart and timing of the LIF module with noise}
\end{figure}
 
 Fig. 5(a) shows the update of the digital LIF in the KWN mode comprising 3 pipelined processes of leak, update and compare. Since only K non-zero $Z_j$ are received from the crossbar, it is possible to introduce large errors in spike times of neurons close to the threshold $V_{th1}$. Hence, a SNL is maintained such that the $V_{mem}$ of these neurons satisfy $V_{th2}$$<V_{mem}$$<$$V_{th1}$. Further, a Pseudo-Random Binary Sequence (PRBS) is used to generate the noise term n(t) in (1) which enables neurons in SNL to probabilistically fire a spike. SNL and noise addition result in 0.5-0.6\% accuracy improvements in two datasets in Fig. 5(b). 
 \begin{figure}[t]
  \centering
  \includegraphics[width=0.95\linewidth]{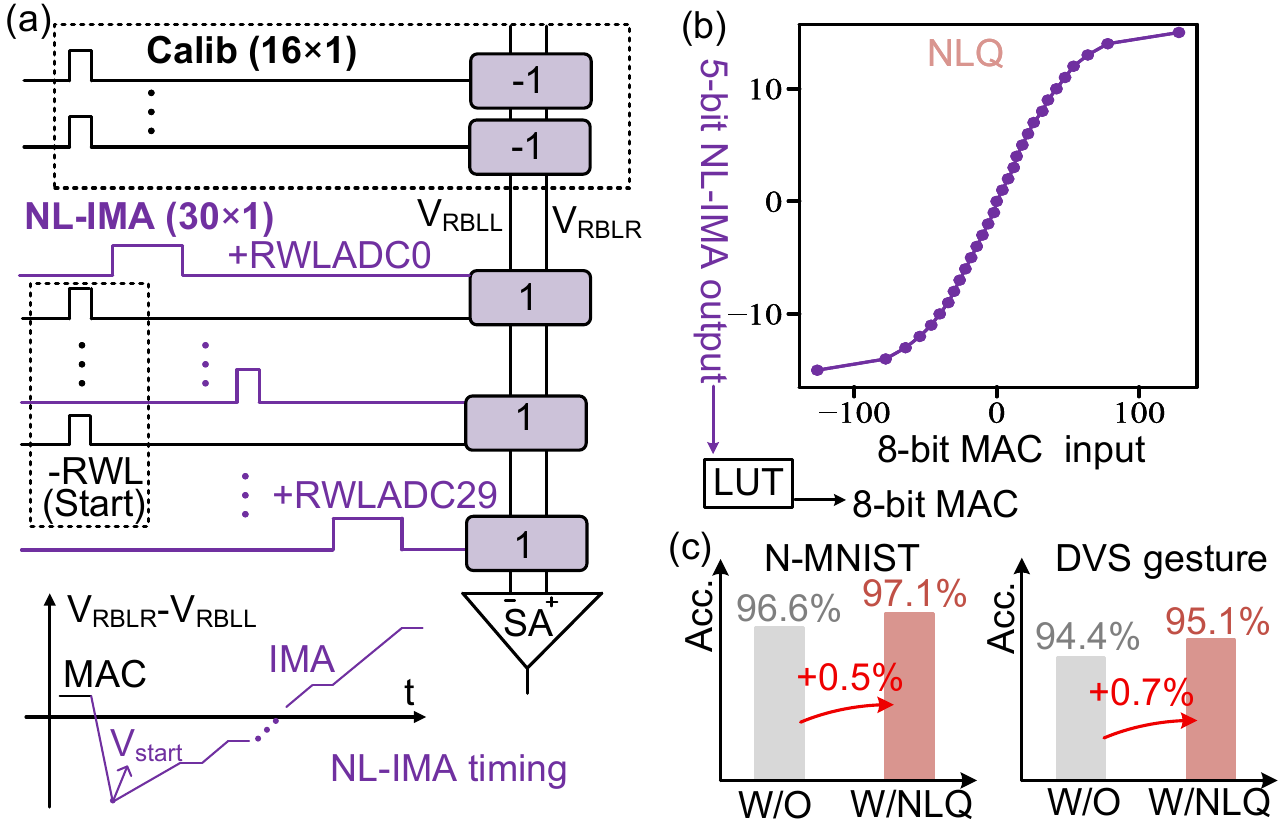}
  \caption{(a) NL-IMA timing with calibration. (b)  NL-IMA for NLQ. (c) Accuracy improvement with NLQ.}
  \label{NL-IMA for NL quantization}
\end{figure}

The ramp IMA can be operated in NL mode by using variable pulse width for each row (Fig. 6(a), blue +RWLADC). This can be used to create an NL quantization of the MAC as shown in Fig. 6(b) in KWN mode, which helps to reduce ADC resolution (e.g., 5-bit ADC for 8-bit range) and thus energy/latency as well. These NL values are helpful for increasing accuracy (0.5-0.7\%) for two datasets in Fig. 6(c) when used in training. In KWN mode, they are mapped back to 8-bit values using a Look-up table (LUT).  

In NLD mode, the reconfigurable NL-IMA approximates various NL activations ($f()$ in (2)) by modulating the pulse width of each quantization step as shown in Fig. 6(a), effectively enhancing neuronal learning capabilities across diverse neuromorphic tasks.

\section{Measurement Results }
 \begin{figure}[t]
  \centering
  \includegraphics[width=0.95\linewidth]{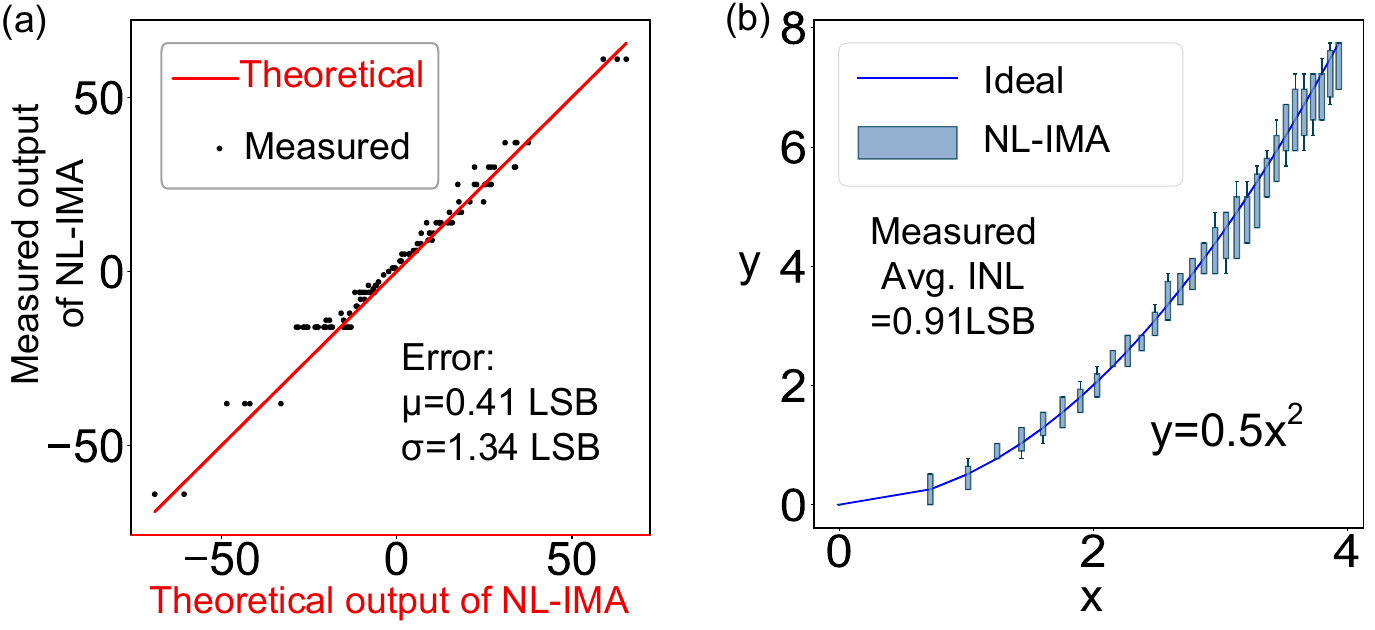}
  \caption{Measurement results of NL-IMA for (a) NLQ and (b) NL activation.}
  \label{Measurement results of NL-IMA for NLQ and NL}
\end{figure}
NeuDW-CIM is fabricated in 65 nm CMOS, and the die micrograph is shown in Fig. 9(c), which occupies 920 $\mu$m×1200 $\mu$m with an active area of 0.513 mm\textsuperscript{2} (570 $\mu$m×900 $\mu$m). 

Fig. 7(a) compares the measured outputs of NL-IMA for NLQ against theoretical values. The measured data points exhibit a strong correlation with the ideal transfer curve, showing a minor mean error ($\mu$) of 0.41 LSB and a standard deviation ($\sigma$) of 1.34 LSB. Furthermore, Fig. 7(b) demonstrates the reconfigurability of NL-IMA in NLD mode by implementing a quadratic activation function ($y=0.5x^2$). The NL-IMA output closely tracks the ideal value with a measured average integral non-linearity (INL) of only 0.91 LSB. 
 \begin{figure}[t]
  \centering
  \includegraphics[width=\linewidth]{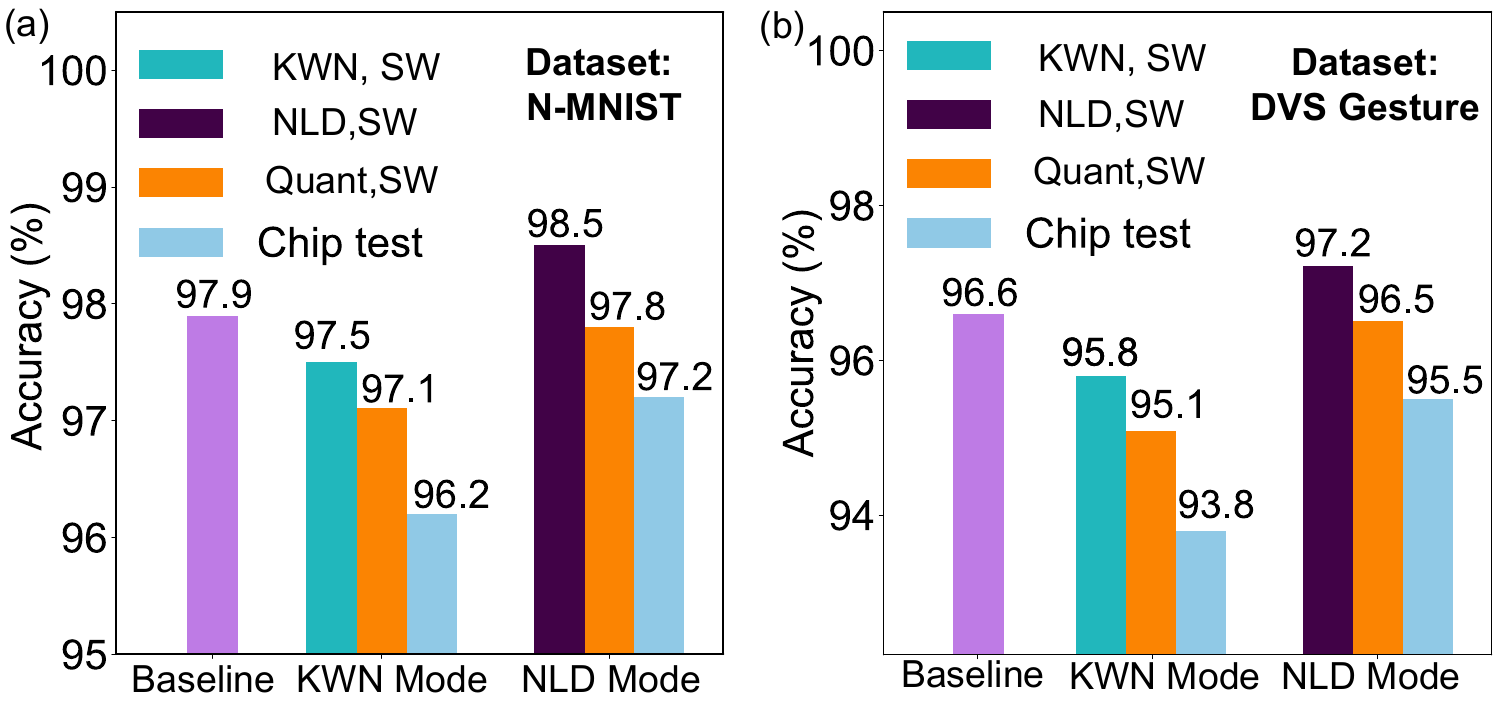}
  \caption{Accuracy under KWN, NLD modes using 3-bit weight and 5-bit NL-IMA. (a) N-MNIST dataset. (b) DVS Gesture dataset.}
  \label{Accuracy under KWNNLD modes with different datasets}
\end{figure}

Fig. 8 shows accuracies on the N-MNIST and DVS Gesture datasets with 3-bit weight and 5-bit NL-IMA. The NLD mode achieves high measured accuracies of 97.2\% and 95.5\% on two benchmark datasets, respectively. Meanwhile, the KWN mode also attains competitive measured results (96.2\% and 93.8\%, respectively) by incorporating the proposed SNL and NLQ methods.

Fig. 9(a) depicts the energy breakdown of the two modes at VDD = 0.7 V. Measured EEs of K=3 (N-MNIST) and K=12 (DVS Gesture) under different VDD are shown in Fig. 9(b). At VDD = 0.7 V and K=3, NeuDW-CIM achieves the best EE of 0.8 pJ/SOP in KWN mode. The proposed reconfigurable IMA  supports both linear and NL modes with a minimal area overhead of only 12\%.
\begin{figure}[t]
  \centering
  \includegraphics[width=0.92\linewidth]{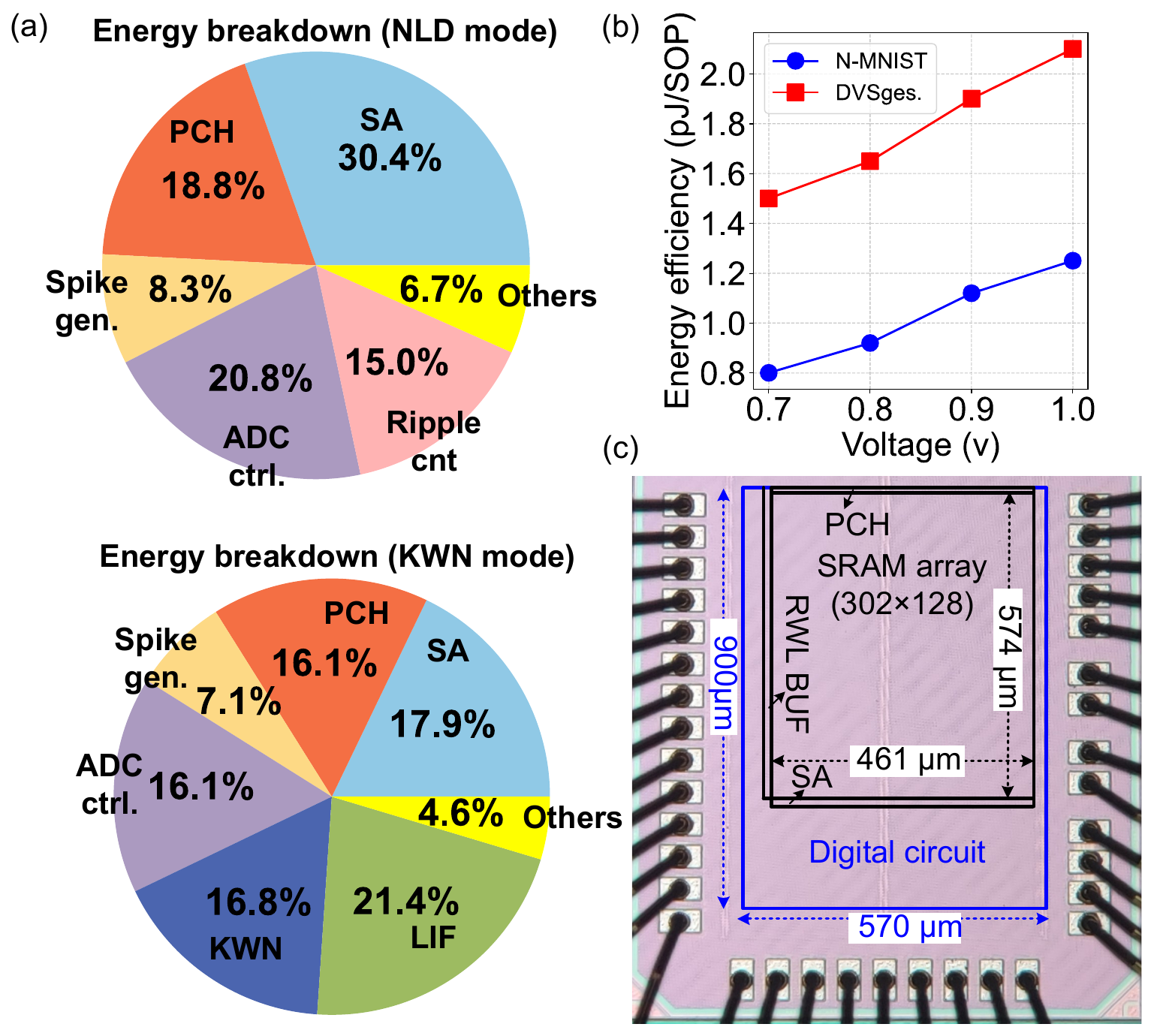}
  \caption{(a) Energy consumption breakdown for NLD and KWN modes. (b) Energy efficiency across various supply voltages.  (c) Die micrograph.}
  \label{Hardware performance based on measurement}
\end{figure}  

 Table I compares NeuDW-CIM with other recent works \cite{ESSRCICsharma202565nm,1zhang202322,4liu202430,10fu2025neuc}. This work achieves better accuracies and  EEs on multiple datasets (97.2\% on N-MNIST at 1.8 pJ/SOP, 95.5\% on DVS Gesture at 2.3 pJ/SOP) in NLD mode. The early stopping feature of the KWN mode enables higher EEs of 0.8 pJ/SOP (1.6× improvement over the SOTA result 1.3 pJ/SOP in \cite{10fu2025neuc}) for K=3 (N-MNIST) and 1.5 pJ/SOP for K=12 (DVS Gesture). While a traditional digital LIF requires serial updates for the entire 128 neurons, our KWN-based sparse update reduces the LIF latency by 10$\times$, as only the K=12 identified winners necessitate $V_{mem}$ update. Furthermore, the additional KWN control logic for early stopping accounts for merely $16.8\%$ of the total power as shown in Fig. 9(a). To show versatility, we tested the macro on a ternary input spike detection dataset (Quiroga \cite{11akhoundi2025scalable}) and achieved 2.1 pJ/SOP EE, 96.1\% accuracy in NLD mode. 

\begin{table}[t!]
  \centering
  \caption{Comparison with the state-of-the-art neuromorphic CIM
  \label{tab:bitcell_comparison}}
  \includegraphics[width=\columnwidth]{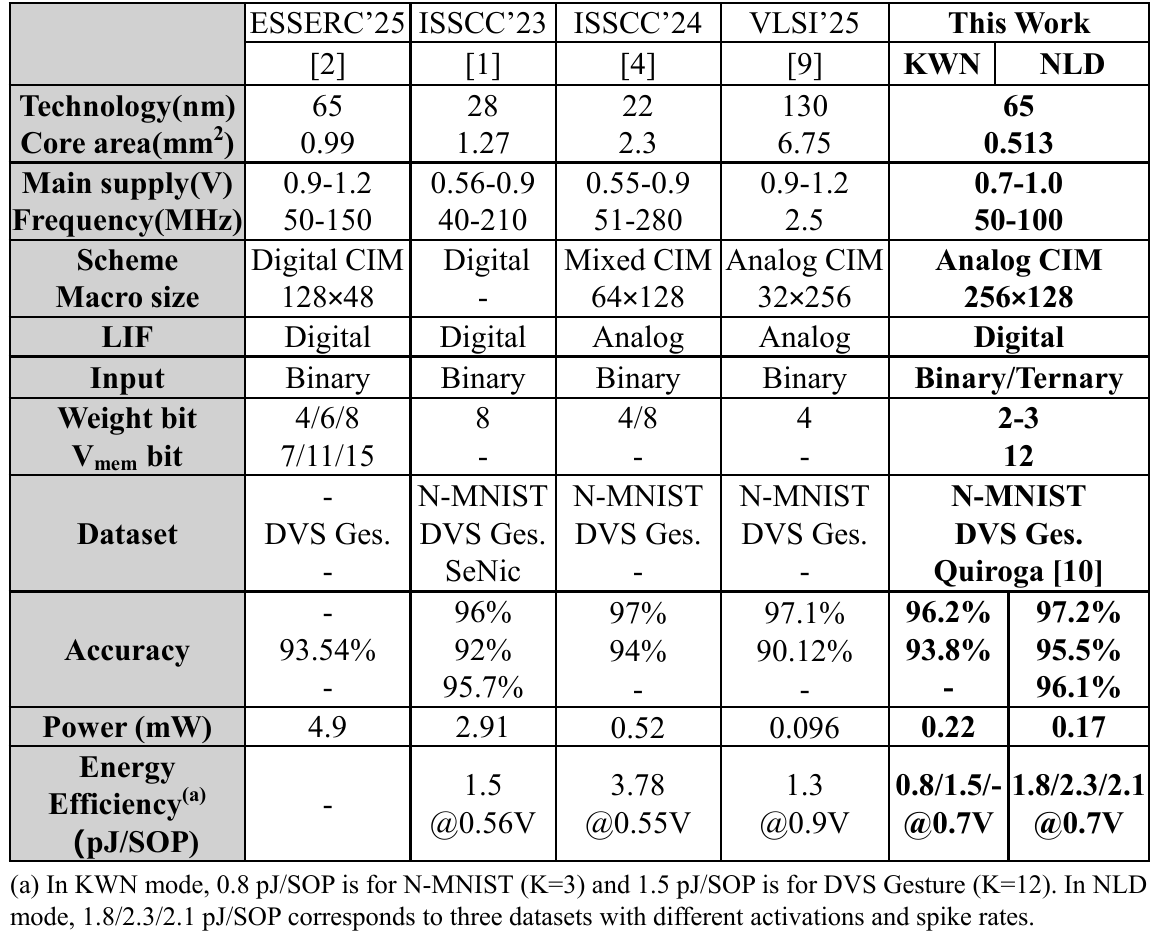} 
\end{table}

\section{Conclusion}
In this work, NeuDW-CIM offers a power-efficient solution for neuromorphic applications via a reconfigurable NL-IMA supporting NLD and KWN modes. Utilizing a multi-$V_{DD}$ twin 9T bit-cell for ternary operations and an early-stopping mechanism, the macro minimizes area overhead for implementing multi-bit weight while reducing ADC and digital LIF latencies by 30\% and 10$\times$. NLQ and SNL are utilized to improve inference accuracy in KWN mode. Ultimately, the 65-nm prototype achieves high energy efficiency and inference accuracy on three datasets. 





\bibliography{IEEEref}

@inproceedings{1zhang202322,
  title={{anp-i: A 28nm 1.5 pJ/sop asynchronous spiking neural network processor enabling sub-o. 1 $\mu$J/sample on-chip learning for edge-AI applications}},
  author={Zhang, Jilin and others},
  booktitle={2023 IEEE International Solid-State Circuits Conference (ISSCC)},
  pages={21--23},
  year={2023},
  organization={IEEE}
}

@inproceedings{3niwa20232,
  title={{A 2.97 $\mu$m-pitch event-based vision sensor with shared pixel front-end circuitry and low-noise intensity readout mode}},
  author={Niwa, Atsumi and others},
  booktitle={2023 IEEE International Solid-State Circuits Conference (ISSCC)},
  pages={4--6},
  year={2023},
  organization={IEEE}
}

@inproceedings{4liu202430,
  title={{a 22nm 0.26 nw/synapse spike-driven spiking neural network processing unit using time-step-first dataflow and sparsity-adaptive in-memory computing}},
  author={Liu, Ying and others},
  booktitle={2024 IEEE International Solid-State Circuits Conference (ISSCC)},
  volume={67},
  pages={484--486},
  year={2024},
  organization={IEEE}
}

@article{6kim2023neuro,
  title={{Neuro-CIM: ADC-less neuromorphic computing-in-memory processor with operation gating/stopping and digital--analog networks}},
  author={Kim, Sangyeob and others},
  journal={IEEE Journal of Solid-State Circuits},
  volume={58},
  number={10},
  pages={2931--2945},
  year={2023},
  publisher={IEEE}
}

@article{7dong2025topkima,
  title={{Topkima-Former: Low-Energy, Low-Latency Inference for Transformers Using Top-k In-Memory ADC}},
  author={Dong, Shuai and others},
  journal={IEEE Transactions on Circuits and Systems I: Regular Papers},
  year={2025},
  publisher={IEEE}
}

@article{8yu202265,
  title={{A 65-nm 8T SRAM compute-in-memory macro with column ADCs for processing neural networks}},
  author={Yu, Chengshuo and others},
  journal={IEEE Journal of Solid-State Circuits},
  volume={57},
  number={11},
  pages={3466--3476},
  year={2022},
  publisher={IEEE}
}

@inproceedings{10fu2025neuc,
  title={{NeuC-CIM: A 1.3 pJ/SOP Neuromorphic Charge-Domain Compute-in-Memory Macro for Spiking Neural Network}},
  author={Fu, Haotian and others},
  booktitle={2025 Symposium on VLSI Technology and Circuits (VLSI Technology and Circuits)},
  pages={1--3},
  year={2025},
  organization={IEEE}
}

@article{11akhoundi2025scalable,
  title={{A Scalable 1024-Channel Ultra-Low-Power Spike Sorting Chip With Event-Driven Detection and Spatial Clustering}},
  author={Akhoundi, Arash and others},
  journal={IEEE Journal of Solid-State Circuits},
  year={2025},
  publisher={IEEE}
}

@inproceedings{ESSRCICsharma202565nm,
  title={{A 65nm 5 TOPS/W Digital CIM Accelerator with Reconfigurable Precision and Temporal Pipelining for Spiking Neural Networks}},
  author={Sharma, Deepika and others},
  booktitle={2025 IEEE European Solid-State Electronics Research Conference (ESSERC)},
  pages={5--8},
  year={2025},
  organization={IEEE}
}

@inproceedings{ESSRICchoi2024ar,
  title={{AR-CIM: A 460.22 TOPS/W SRAM-based Analog Reconfigurable Computing-in-Memory Macro with 1/2/4/8-Bit Variable Precision}},
  author={Choi, Byeongseon and others},
  booktitle={2024 IEEE European Solid-State Electronics Research Conference (ESSERC)},
  pages={361--364},
  year={2024},
  organization={IEEE}
}
\bibliographystyle{IEEEtran}

\end{document}